\documentclass[aps,prd,twocolumn,groupedaddress,showpacs,showkeys]{revtex4-1}
\usepackage{graphicx}
\usepackage{dcolumn}
\usepackage{bm}
\usepackage{amssymb,amsmath,latexsym,amsfonts}
\begin{document}


\title{Bogoliubov space of a Bose--Einstein condensate and quantum spacetime fluctuations}

\author{J. I. Rivas}
 \email{jirs@xanum.uam.mx}\affiliation{Departamento de F\'{\i}sica,
 Universidad Aut\'onoma Metropolitana--Iztapalapa\\
 Apartado Postal 55--534, C.P. 09340, M\'exico, D.F., M\'exico.}

 \author{A. Camacho}
 \email{acq@xanum.uam.mx} \affiliation{Departamento de F\'{\i}sica,
 Universidad Aut\'onoma Metropolitana--Iztapalapa\\
 Apartado Postal 55--534, C.P. 09340, M\'exico, D.F., M\'exico.}

 \author{E. G\"okl\"u}
 \email{cbi920000370@xanum.uam.mx} \affiliation{Departamento de F\'{\i}sica,
 Universidad Aut\'onoma Metropolitana--Iztapalapa\\
 Apartado Postal 55--534, C.P. 09340, M\'exico, D.F., M\'exico.}


\date{\today}

\begin{abstract}
In the present work we consider the role that metric fluctuations
could have upon the properties of a Bose--Einstein condensate. In
particular we consider the Bogoliubov space associated to it and
show that there are, at least, two independent ways in which the
average size of these metric fluctuations could be, experimentally,
determined. Indeed, we prove that the pressure and the speed of
sound of the ground state define an expression allowing us to
determine the average size of these fluctuations. Afterwards, an
interferometric experiment involving Bogoliubov excitations of the
condensate and the pressure (or the speed of sound of the ground
state) provides a second and independent way in which this average
size could be determined, experimentally.
\end{abstract}

\maketitle

\section{Introduction}

The mathematical and physical difficulties plaguing all theoretical
models behind a quantum theory of gravity \cite{kop04, beck1} have
spurred the so--called quantum gravity phenomenology, a topic that
can be defined as the observational and experimental search for
deviations from Einstein's General Relativity, or from quantum field
theory. It embodies deformed versions of the dispersion relation
\cite{Giov5, Giov15}, or deviations from the $1/r$--potential and
violations of the equivalence principle
\cite{acclamhm,FischbachTalmadge99}. Of course, these aforementioned
cases do not exhaust the extant possibilities. The use of cold
atoms, either bosonic or fermionic, in this context is a point
already considered \cite{Coll1, Coll2}. In particular the
possibility of constraining the energy--momentum relation resorting
to cold atoms has already shown us that this kind of systems could
open up new landscapes in the context of gravitational physics
\cite{Giovanni34}.

In the present work we consider the role that metric fluctuations
could have upon the properties of a Bose--Einstein condensate. In
particular we address the issue of conformal metric fluctuations.
Let us explain, briefly, the meaning of this last sentence. In this
context the main idea corresponds to a Minkowskian background and in
addition small spacetime fluctuations are also present, and they are
a consequence of some quantum gravity scenario. One of the
assumptions in this approach is related to the fact that these
spacetime fluctuations emerge as classical fluctuations of the
background metric. There are several possibilities around the type
of fluctuations that can be considered \cite{Crowell}. among the
huge spectrum of possibilities we may find the so--called conformal
fluctuations \cite {Ertan1, Ertan2}. They can be considered,
mathematically, the simplest case and entail a redefinition of the
inertial mass. In other words, the average size of this kind of
fluctuations, say $\gamma$, appears in the corresponding motion
equations in the form $m/(1+\gamma)$, where $m$ denotes the usual
inertial mass (when no fluctuations are present). Clearly, an
experimental outcome detects $m/(1+\gamma)$, and not $m$ or $\gamma$
separately. Of course, a possible objection to the analysis of
conformal fluctuations could be related to its simplicity, which in
physical terms implies a redefinition of the inertial mass. In this
work we will show that even the simplest case, i.e., conformal
fluctuations, is endowed with a richness that leads us to detectable
effects.

The idea in the present work is related to the possibility of, by
means of two or more experimental proposals, deducing, separately,
$m$ and $\gamma$. It will be proved that, the Bogoliubov space of a
Bose--Einstein offers this option. Indeed, we consider the
Bogoliubov space associated to a Bose--Einstein condensate and show
that there are, at least, two independent ways in which the average
size of these metric fluctuations can be, experimentally determined.
Firstly, we consider the many--particle Hamiltonian of a bosonic
gas, immersed in a homogeneous gravitational field, and, in
addition, the effects of the metric fluctuations upon the inertial
mass will be introduced. It will be shown that the pressure and
speed of sound of the ground state of the Bogoliubov space of the
condensate allow us to put forward and experiment which, in
principle, determines the average size of these fluctuations. As a
by product the value of $m$, i.e., the bare mass, can also be
obtained. Secondly, we analyze an interferometric experiment,
resorting to Bogoliubov excitations, and deduce the phase shift
induced by the gravitational field and the metric fluctuations. It
will be proved that this gravity--induced phase shift together with
the pressure (or the speed of sound) of the ground state of the
Bogoliubov space imply two expressions which determine the size of
these metric fluctuations, and also of $m$. In order to understand
better this argument, let us denote by $v_s$ the speed of sound of
the ground state of the Bogoliubov space, and by $\Delta \phi$ the
aforementioned gravity--induced phase shift of two Bogoliubov
excitations. It will be shown that, roughly said, $m/(1+\gamma) =
f_1(v_s, a, N, V, g)$, whereas $m/\sqrt{(1+\gamma)} =
f_2(\Delta\phi, a, N, V, g)$. In this last two expression $f_1$ and
$f_2$ are two functions (deduced in this work), and $N, V, a$, and
$g$, denote the number of particles, volume of the container of the
bosonic gas, scattering length, and, finally, acceleration of
gravity, respectively. Clearly, $f_2/f_1 = \sqrt{1+\gamma}$. In
other words, the experimental deduction of $f_1$ and $f_2$ leads us
to the determination of a characteristic of this metric
fluctuations, namely, $\gamma = (f_2/f_1)^2 -1$. In a similar way we
may find an expression for $m$.
\bigskip
\bigskip

\section{Metric fluctuations and a weakly interacting Bose gas}
\bigskip

\subsection{Metric fluctuations}
\bigskip

Now we, succinctly, address the issue of metric fluctuations. In
this context a spacetime is present, the one can be regarded as a
classical background on which classical fluctuations exist.
\cite{Ertan1}. We suppose they are a consequence of a quantum theory
of gravity where the microscopic structure of spacetime exhibits
quantum fluctuations. These spacetime fluctuations modify our
current physics. Indeed, they entail a change of the motion
equations. In this general scheme these fluctuations imply a
modified Schr\"odinger equation, in which the Laplacian operator
becomes $(\delta^{ij} + \gamma^{ij})\partial_{i}\partial_{j}$, where
$\gamma^{ij}>0$. \cite{Ertan1, Ertan2}. The simplest case involves
the so--called conformal fluctuations defined by the following
conditions: $\gamma^{ij} = 0$, if $i\not =j$, whereas $\gamma^{xx} =
\gamma^{yy} = \gamma^{zz} = \gamma$. We may rephrase this last
assertion stating that the idea of conformal fluctuations is
depicted by matrices proportional to the unit matrix. Additionally,
these fluctuations can be comprehended (at least partially) as
redefinitions of the inertial mass \cite{Ertan2}. The new inertial
mass for an atom ($m^{eff}$) is now given by

\begin{equation}
m^{eff}=m\Bigl(1+\gamma\Bigr)^{-1}.\label{ele1}
\end{equation}

The parameter $\gamma$ encodes the information concerning the metric
fluctuations and its value depends on the particle species under
consideration. Clearly, they have to be considered very small
(otherwise its existence would be a proved experimental fact). In
addition, we do assume that it has a particular sign, namely, it has
to be positive \cite{Ertan2}. Additionally, ($\gamma$) depends upon
the type of particle \cite{Ertan1, Ertan2}.

This last expression contains an experimental hurdle. Indeed, notice
that a kinematical experiment detects the relation
$m\Bigl(1+\gamma\Bigr)^{-1}$, but not $m$ or $\gamma$, separately.
At this point we pose the main question that will be addressed in
this work: Could $m$ and $\gamma$ be measured separately?
\bigskip

\subsection{Weakly interacting gas}
\bigskip

As mentioned before, the main idea in this work involves an
experimental proposal for the detection of the {\it bare inertial
mass}, here denoted by $m$, and the average size of our conformal
fluctuations, the parameter $\gamma$. Indeed, the Laplacian operator
becomes, under this condition \cite{Ertan1, Ertan2}.

\begin{equation}
\Delta_{\gamma}=(\delta^{ij} +
\gamma^{ij})\partial_{i}\partial_{j}.\label{lap1}
\end{equation}

This modification will be considered in the $N$--body Hamiltonian
operator (assuming that the gas is so dilute that only the two--body
interaction potential is required \cite{Pathria}). Our model will be
a Bose--Einstein gas enclosed in a container of volume $V$,
particles of the gas are atoms with passive gravitational mass $m$
and located at a height $l$ with respect to the Earth´s surface. The
interaction between two particles will be assumed to be dominated by
$s$--scattering, i.e., the temperature of the system is very low
($ka<<1$, where $\vec{k}$ and $a$ are the wave vector and the
scattering length, respectively) \cite{Pitaevski}. This entails the
following Hamiltonian for the $N$--body system.

\begin{eqnarray}
\hat{H} =
\sum_{\vec{k}=0}\frac{\hbar^2k^2}{2m}(1+\gamma)\hat{a}_{\vec{k}}^{\dagger}\hat{a}_{\vec{k}}\nonumber\\
+\frac{U_0}{2V}\sum_{\vec{k}=0}\sum_{\vec{p}=0}\sum_{\vec{q}=0}\hat{a}_{\vec{p}}^{\dagger}\hat{a}_{\vec{q}}^{\dagger}
 \hat{a}_{\vec{p}+\vec{k}}\hat{a}_{\vec{q}-\vec{k}} \nonumber\\
 +
 \sum_{\vec{k}=0}mgl\hat{a}_{\vec{k}}^{\dagger}\hat{a}_{\vec{k}}, \label{Ham1}
\end{eqnarray}

\begin{equation}
U_0=\frac{4\pi a\hbar^2}{m}(1+\gamma).\label{Add1}
\end{equation}

Demanding $\gamma=0$ we recover the usual result \cite{Pethick}.
These operators ($\hat{a}_{\vec{k}}$ and
$\hat{a}_{\vec{k}}^{^{\dagger}}$) are bosonic creation and
annihilation operators, and fulfill the usual Bose commutator
relations. Very close to the temperature $T=0$, the second term in
this Hamiltonian becomes \cite{Pethick}

\begin{eqnarray}
\sum_{\vec{k}=0}\sum_{\vec{p}=0}\sum_{\vec{q}=0}\hat{a}_{\vec{p}}^{\dagger}\hat{a}_{\vec{q}}^{\dagger}
 \hat{a}_{\vec{p}+\vec{k}}\hat{a}_{\vec{q}-\vec{k}} = N^2+2N\sum_{\vec{k}\not=0}\hat{a}_{\vec{k}}
 ^{\dagger}\hat{a}_{\vec{k}}\nonumber\\
 + N\sum_{\vec{k}\not=0}\Big(\hat{a}_{\vec{k}}^{\dagger}\hat{a}_{-\vec{k}}^{\dagger}
+ \hat{a}_{\vec{k}}\hat{a}_{-\vec{k}}\Big). \label{Add2}
\end{eqnarray}

With this approximation the $N$--body Hamiltonian has the following
structure

\begin{eqnarray}
\hat{H} = \frac{U_0N^2}{2V}+mglN+\nonumber\\
\sum_{\vec{k}\not=0}\Bigl[\frac{\hbar^2k^2}{2m}(1+\gamma)+mgl+\frac{U_0N}{V}\Bigr]\hat{a}_{\vec{k}}
^{\dagger}\hat{a}_{\vec{k}}\nonumber\\
+N\frac{U_0}{2V}\Bigl[\hat{a}_{\vec{k}}^{\dagger}\hat{a}_{-\vec{k}}^{\dagger}
+ \hat{a}_{\vec{k}}\hat{a}_{-\vec{k}}\Bigr]. \label{Ham2}
\end{eqnarray}

This Hamiltonian can be diagonalized introducing the
Bogoliubov transformations \cite{Pitaevski}

\begin{equation}
\hat{b}_{\vec{k}}=
\frac{1}{\sqrt{1-\alpha_k^2}}\Bigl[\hat{a}_{\vec{k}} +
\alpha_k\hat{a}_{-\vec{k}}^{\dagger}\Bigr],\label{Bog1}
\end{equation}

\begin{equation}
\hat{b}_{\vec{k}}^{\dagger}=
\frac{1}{\sqrt{1-\alpha_k^2}}\Bigl[\hat{a}_{\vec{k}}^{\dagger} +
\alpha_k\hat{a}_{-\vec{k}}\Bigr].\label{Bog2}
\end{equation}

In this last expression the following definitions have been
introduced

\begin{equation} \epsilon_k =
\frac{\hbar^2k^2}{2m}(1+\gamma)+mgl,\label{Add3}
\end{equation}

\begin{equation}
\alpha_k = 1+ \frac{V\epsilon_k}{U_0N}
-\sqrt{\frac{V\epsilon_k}{U_0N}}\sqrt{2+
\frac{V\epsilon_k}{U_0N}}.\label{Add33}
\end{equation}

They fulfill the same algebra related to $\hat{a}_{\vec{k}}$ and
$\hat{a}_{\vec{k}}^{\dagger}$, i.e., they are also bosonic
operators. The final form for our Hamiltonian is

\begin{eqnarray}
\hat{H} = \frac{U_0N^2}{2V} + mglN \nonumber\\
+
\sum_{\vec{k}\not=0}\Bigl\{\sqrt{\epsilon_k(\epsilon_k+\frac{2U_0N}{V})}
\hat{b}_{\vec{k}}^{\dagger}\hat{b}_{\vec{k}}\nonumber\\
-\frac{1}{2}\Bigl[\frac{U_0N}{V} +\epsilon_k
-\sqrt{\epsilon_k(\epsilon_k+\frac{2U_0N}{V})}\Bigr]\Bigr\}.\label{Ham3}
\end{eqnarray}

The last summation diverges, a result already known \cite{Gribakin,
Ueda}, and this divergence disappears introducing the so--called
pseudo--potential method, which implies that we must perform the
following substitution \cite{Ueda}

\begin{eqnarray}
 &-\frac{1}{2}\Bigl[\frac{U_0N}{V} +\epsilon_k -\sqrt{\epsilon_k(\epsilon_k+\frac{2U_0N}{V})}\Bigr]&\rightarrow\nonumber \\
 &-\frac{1}{2}\Bigl[\frac{U_0N}{V}
 +\epsilon_k -\sqrt{\epsilon_k(\epsilon_k+
 \frac{2U_0N}{V})}-\frac{1}{2\epsilon_k}\bigl(\frac{U_0N}{V}\bigr)^2\Bigr]&.\label{Psedo1}
\end{eqnarray}

Finally, this last summation will be approximated by an integral. It
is noteworthy to mention that the original expression has as lower
limit the condition $k\not=0$, which implies that the integral has
as lower limit the value $mgl$. In other words,

\begin{eqnarray}
&-&\frac{1}{2}\sum_{\vec{k}\neq 0}\Bigl[\frac{U_0N}{V} +\epsilon_k
-\sqrt{\epsilon_k(\epsilon_k+
 \frac{2U_0N}{V})}-\frac{1}{2\epsilon_k}\bigl(\frac{U_0N}{V}\bigr)^2\Bigr]
 \nonumber\\
 &=& -\frac{\hbar^2V}{8m\pi^2}\bigl(\frac{8\pi aN}{V}\bigr)^{5/2}\int_{mgl}^{\infty}f(x)dx. \label{Psedo2}
\end{eqnarray}

In this last expression we have that

\begin{eqnarray}
f(x) =
x^2\Bigl[1+x^2-x\sqrt{2+x^2}-\frac{1}{2x^2}\Bigr].\label{Psedo3}
\end{eqnarray}

In this last expression we have that

\begin{eqnarray}
x = \sqrt{\frac{\epsilon_kV}{U_0N}}.\label{Psedo33}
\end{eqnarray}

With these conditions we deduce the final structure of the $N$--body
Hamiltonian

\begin{equation}
\hat{H} = E_0 +
\sum_{\vec{k}\not=0}E_k\hat{b}_{\vec{k}}^{\dagger}\hat{b}_{\vec{k}}.\label{Ham4}
\end{equation}

In this last expression $E_0$ denotes the energy of the ground state
of the corresponding Bogoliubov space \cite{Pitaevski}.

\begin{eqnarray}
E_0 = \frac{2\pi a\hbar^2N^2(1+\gamma)}{mV}\Bigl[1 +
\frac{128}{15}\sqrt{\frac{a^3N}{V\pi}}\nonumber\\
\Bigl(1-\frac{15} {16\sqrt{2}}\sqrt{\frac{m^2glV}{4\pi
a\hbar^2N(1+\gamma)}}\Bigr)\Bigr] +Nmgl. \label{E0}
\end{eqnarray}

On the other hand, we have that the energy of the Bogoliubov
excitations ($E_k$) is given by \cite{Pitaevski}

\begin{equation}
E_k = \sqrt{\epsilon_k\Bigl(\epsilon_k+
 \frac{2U_0N}{V}\Bigr)}. \label{Ek}
\end{equation}

Concerning (\ref{Ham4}), if we impose the condition of vanishing
gravitational constant, i.e. $g=0$, then we recover the usual
Hamiltonian \cite{Ueda}.
\bigskip

\subsection{Speed of sound and pressure of the ground state}
\bigskip

The pressure ($P_0 = -\frac{\partial E_0}{\partial V}$) and speed of
sound ($v_s =\sqrt{-\frac{{V}^{2}}{N m}\frac{\partial P_0}{\partial
V}}$) associated to the ground state of the Bogoliubov space become,
respectively

\begin{eqnarray}
P_0 = \frac{2\pi a\hbar^2N^2(1+\gamma)}{mV^2}\Bigl[1 +
\frac{192}{15}\sqrt{\frac{a^3N}{V\pi}}\nonumber\\
\Bigl(1-\frac{5} {8\sqrt{2}}\sqrt{\frac{m^2glV}{4\pi
a\hbar^2N(1+\gamma)}}\Bigr)\Bigr], \label{Pre0}
\end{eqnarray}

\begin{eqnarray}
v^2_s = \frac{4\pi a\hbar^2N(1+\gamma)^2}{m^2V}\Bigl[1 +
16\sqrt{\frac{a^3N}{V\pi}}\nonumber\\
\Bigl(1-\frac{1}{2\sqrt{2}}\sqrt{\frac{m^2glV}{4\pi
a\hbar^2N(1+\gamma)}}\Bigr)\Bigr]. \label{Sp0}
\end{eqnarray}

A fleeting glimpse at these last expressions tells us that they
depend upon the {\it bare inertial mass} and on the {\it size} of
the fluctuations, i.e, upon $m$ and $\gamma$, respectively.

Notice that they imply the following relation
\begin{eqnarray}
\frac{m}{\sqrt{1+\gamma}}=\frac{{v}_{s}^{2}\pi
a{\hbar}^{2}{N}^{3}{(1+\alpha)}^{2}-{P}_{0}^{2}{V}^{3}\Bigl(1+\frac{5}{4}\alpha\Bigr)}{2\alpha{v}_{s}^{2}\pi
a{\hbar}^{2}\beta(1+\alpha)-{P}_{0}^{2}{V}^{3}\alpha\beta}\label{rel1},
\end{eqnarray}

\begin{eqnarray}
\alpha=\frac{192}{15}\sqrt{\frac{{a}^{3}N}{V\pi}};\qquad
\beta=\frac{5}{8\sqrt{2}}\sqrt{\frac{glV}{4\pi a{\hbar}^{2}N}}.
\end{eqnarray}

Clearly, it does not contain the usual relation, namely,
$m(1+\gamma)^{-1}$. Consider the kinematical relation associated to
$U_0$ (see (\ref{Add1})). If $F$ denotes the right--hand side of
(\ref{rel1}), then we obtain

\begin{equation}
\sqrt{(1+\gamma)} =  \frac{FU_0}{4\pi a\hbar^2}, \label{rel2}
\end{equation}

\begin{equation}
m =  \frac{F^2U_0}{4\pi a\hbar^2}. \label{rel3}
\end{equation}

These two last expression allows us to deduce $m$ and $\gamma$,
separately. The right--hand side of them involves parameters which
can be detected experimentally.
\bigskip

\subsection{Bogoliubov excitations and metric fluctuations}
\bigskip

\begin{figure}[h]
\begin{center}
\includegraphics[scale=1.2]{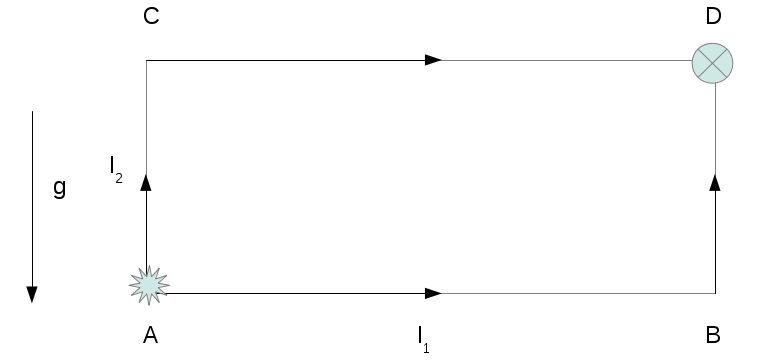}
\caption{\label{figure1} The interferometer with arm lengths $l_1$
and $l_2$. The source for the Bose-Einstein-Condensates is located
at point A, where it is coherently split into two sub-condensates
which travel along different paths. Finally, due to the different
times of flight, the phase shift $\Delta \phi$ can be detected at
the detector at point D by means of the interference pattern.}
\end{center}
\end{figure}

We now present a second manner in which $m$ and $\gamma$ could be
detected separately. Here we will resort to the properties of
Bogoliubov excitations of the condensate. It is already known
\cite{Pitaevski} that, even at $T=0$, the presence of two--body
interactions entail the existence of excitations in the condensate
\cite{Pethick, Ueda}, whose energy is given by (\ref{Ek}). At this
point we consider two Bogoliubov excitations located, initially, at
point (A) (see figure \ref{figure1}), and whose wave vector fulfills
the following condition

\begin{equation}
\frac{\hbar^2k^2}{2m}(1+\gamma)+mgl>\frac{2U_0N}{V}. \label{Int1}
\end{equation}

Then, we have, approximately

\begin{equation}
E_k=\frac{\hbar^2k^2}{2m}(1+\gamma)+mgl +\frac{4\pi
a\hbar^2N(1+\gamma)}{mV}. \label{Int2}
\end{equation}

These conditions allow us to consider an interferometric proposal
along the lines of a semi--classical approximation. In other words,
we consider an experiment similar to the Colella--Overhauser--Werner
idea \cite{COW1, COW2, COW3, COW4}. This last experiment shows the
effects of the gravitational field of the Earth upon the phase shift
of two neutron beams. Here we consider the same kind of proposal but
with a different intention. Indeed, we seek for an extra expression
whose dependence is not of the kind $[m(1+\gamma)^{-1}]^s$, where
$s$ is a real number. We now resort to the WKB approximation in
order to deduce the corresponding gravity--induced phase shift
\cite{Sakurai}.

The time of flight for the beam moving along the path (A)--(B)--(D)
reads

\begin{eqnarray}
^{(1)}t=\frac{ml_1}{(1+\gamma)\hbar
k_0}+\frac{1}{g}\Bigl[\frac{(1+\gamma)}{m}\hbar
k_0\nonumber\\
-\sqrt{\Bigl(\frac{(1+\gamma)}{m}\hbar k_0\Bigr)^2-2gl_2}\Bigr].
\label{Time1}
\end{eqnarray}

Concerning (A)--(C)--(D) the time of flight is

\begin{eqnarray}
^{(2)}t=\frac{1}{g}\Bigl[\frac{(1+\gamma)}{m}\hbar
k_0-\sqrt{\Bigl(\frac{(1+\gamma)}{m}\hbar
k_0\Bigr)^2-2gl_2}\Bigr]\nonumber\\
+ \sqrt{\frac{l^2_1}{\bigl(\frac{(1+\gamma)\hbar k_0}{m}\bigr)^2
-2gl_2}}. \label{Time2}
\end{eqnarray}
\textbf{}

In these last expressions $k_0$ denotes the wave number at point
(A). These expressions are valid for short times of flight or for
large velocities for the motion of the centre of mass of the BEC.
Additionally, we have also assumed that the line passing through
points (B) and (D) is parallel to the direction of the gravitational
field. If this line forms an angle $\theta$ with the gravitational
field, then our expression remains valid but now we have an
effective gravitational acceleration given by $g\cos(\theta)$.
Clearly, with these two flight times we may deduce, easily, the
difference in time of arrival \cite{Sakurai} and, in consequence,
the gravity--induced phase shift, here denoted by $\Delta\phi$.

\begin{equation}
\Delta\phi=\frac{m^2gl_1l_2}{(1+\gamma)\hbar^2k_0}\Bigl[1-\frac{2m^2gl_2}{(1+\gamma)\hbar^2k^2_0}\Bigr].
\label{Phase1}
\end{equation}

Introducing the condition $\gamma=0$ we recover the COW result
\cite{Sakurai}. We now notice that this phase shift contains a
dependence of the kind $m^2/(1+\gamma)$. Taking the predominant term
in the expression for the gravity--induced phase shift and resorting
to our previous results concerning the speed of sound of the ground
state (see (\ref{Sp0})) we may, separately, deduce $m$ and $\gamma$.
Indeed (here $\lambda_0$ denotes the wavelength at (A) divided by
$2\pi$) ,

\begin{equation}
m = \frac{v_s\Delta\phi\hbar^2}{gl_1l_2\lambda_0}\sqrt{\frac{V}{4\pi
aN}}\Bigl[1 - 16\sqrt{\frac{a^3N}{\pi V}}\Bigr]. \label{Res1}
\end{equation}

In a similar way we find that

\begin{equation}
\gamma =
\frac{v^2_s\Delta\phi\hbar^2}{gl_1l_2\lambda_0}\frac{V}{4\pi
aN}\Bigl[1 - 32\sqrt{\frac{a^3N}{\pi V}}\Bigr] -1. \label{Res2}
\end{equation}
\bigskip
\bigskip

In the deduction of these last two expressions we have resorted to
the speed of sound of the ground state, nevertheless, the pressure
could have been used, as well.
\section{Conclusions}
\bigskip

We have introduced into the $N$--body Hamiltonian the effects of
conformal fluctuations of the metric, which are usually associated
to a redefinition of the inertial mass of the involved particles,
i.e., mass appears as $m(1+\gamma)^{-1}$. Two different manners in
which, experimentally, these two parameters could be detected have
been put forward.

At this point we must emphasize the fact that the present work
contains an experimental proposal aimed at the detection of some
properties of quantum fluctuations. Firstly, the deduction of the
pressure and speed of sound of the ground state of the Bogoliubov
space have been calculated. These expressions allow us, when
compared against $U_0$, to achieve our goal. Secondly, an
interferometric proposal, resorting to Bogoliubov excitations has
been analyzed. It has been shown that the gravity--induced phase
shift, along with the pressure (or speed of sound) of the ground
state, renders a second way in which we may, in an experiment,
deduce, separately, $m$ and $\gamma$. It is noteworthy to mention
that this approach can be used in the context of deformed dispersion
relations. Let us explain better this idea. The breakdown of Lorentz
symmetry appears as a consequence in some quantum gravity models
\cite{Giovanni33}. The possibilities that cold atoms offer in this
direction has already been considered, though the role that
Bogoliubov excitations could play in this context has not been
analyzed yet. In other words, the present proposal can be considered
as some kind of complementary study to \cite{Giovanni34}.

Of course, an important point in this context is the feasibility of
the detection of the pressure and speed of sound of the ground state
of a Bose--Einstein condensate. The speed of sound of a
Bose--Einstein condensate, comprising sodium atoms, has already been
measured \cite{Andrews}, though, of course, more experimental work
is needed in this realm. In other words, the present proposal does
not seem to be, in this direction, very far from the present
technology. As an additional bonus related to the present work let
us mention the connection with the Einstein Equivalence Principle
(EEP) \cite{Will93}, namely, the famous ``semicolon goes to coma
rule''. This principle tells us that locally the laws of physics are
the special--relativistic laws. We may rephrase this statement
asserting that locally the gravitational field can be gauged away.
In other words, in a freely falling frame the pressure and speed of
sound, related to the ground state of the Bogoliubov space must be
given by our expression, if $g=0$. Similarly, the gravity--induced
phase shift shall vanish. This kind of experiments are, currently, a
hot topic in gravitational physics \cite{Zoest}, since nowadays it
is possible to create condensates, in a regular basis, under
microgravity. The present idea can also be considered as a proposal
for the use of Bogoliubov fluctuations as an additional tool for
experiments in fundamental physics. This kind of experiments are,
currently, a hot topic in gravitational physics \cite{Zoest}, since
nowadays it is possible to create condensates, in a regular basis,
under microgravity. The present idea can also be considered as a
proposal for the use of Bogoliubov fluctuations as an additional
tool for experiments in fundamental physics.

Finally, a heated debate concerning the role that atom
interferometry plays in the context of gravitational shift
(universality of clock rates) erupted recently \cite{Peters, Wolf}.
Clearly, at least one of these two interpretations has to be wrong.
The aforementioned discussion shall not be neglected since it
addresses an important and fundamental aspect in gravitational
physics. The present proposal puts forward the possibility of
carrying out, with Bogoliubov excitations, a similar experiment. In
other words, it offers the option of an interferometric experiment
without resorting to atoms \cite{Peters, Wolf}, or to neutrons
\cite{Coll1, Coll2}. Maybe this new case could shed some light upon
this debate.
\begin{acknowledgements}
JIRS acknowledges CONACyT grant No. 18176. \\
E. G. acknowledges UAM--I for the grant received.
\end{acknowledgements}

\end{document}